\documentclass[reprint,twocolumn,nofootinbib,prl,aps]{revtex4-1}
\usepackage{amsmath,amssymb,amsfonts,graphicx,color,verbatim,hyperref,fancyref}

\newcommand{\beq}{\begin{equation}}
\newcommand{\eeq}{\end{equation}}
\def\bea{\begin{eqnarray}}
\def\eea{\end{eqnarray}}

\newcommand{\sbreak}{
    \begin{center}
        $\blacklozenge$$\blacklozenge$$\blacklozenge$$\blacklozenge$
    \end{center}
}

\def\({\left(}
\def\){\right)}

\newcommand{\lads}{\ell}
\newcommand{\T}[1]{\langle #1 \rangle_\beta}

\newcommand{\be}{\begin{equation}}
\newcommand{\ee}{\end{equation}}

\begin{document}
\rightline{MIT-CTP/4794}

\title{Lieb-Robinson and the butterfly effect}

\author{Daniel A. Roberts}

\affiliation{Center for Theoretical Physics and Department of Physics,
Massachusetts Institute of Technology, Cambridge, Massachusetts 02139, USA}

\email{drob@mit.edu}

\author{Brian Swingle}

\affiliation{Stanford Institute for Theoretical Physics and Department of Physics,
Stanford University, Stanford, California 94305, USA}

\email{bswingle@stanford.edu}

\begin{abstract}
As experiments are increasingly able to probe the quantum dynamics of systems with many degrees of freedom, it is interesting to probe fundamental bounds on the dynamics of quantum information. We elaborate on the relationship between one such bound---the Lieb-Robinson bound---and the butterfly effect in strongly-coupled quantum systems. The butterfly effect implies the ballistic growth of local operators in time, which can be quantified with the ``butterfly'' velocity $v_B$. Similarly, the Lieb-Robinson velocity places a state independent ballistic upper bound on the size of time evolved operators in non-relativistic lattice models. Here, we argue that $v_B$ is a state-dependent effective Lieb-Robinson velocity. We study the butterfly velocity in a wide variety of quantum field theories using holography and compare with free particle computations to understand the role of strong coupling. We find that, depending on the way length and time scale, $v_B$ acquires a temperature dependence and decreases towards the IR. We also comment on experimental prospects and on the relationship between the butterfly velocity and signaling.
\end{abstract}

\maketitle

In relativistic systems with exact Lorentz symmetry, causality requires that spacelike-separated operators commute. In non-relativistic systems, there is no analogous notion: a local operator $V(0)$ at the origin need not commute with another local operator $W(x,t)$ at position $x$ at a later time $t$, even if the separation is much larger than the elapsed time $|x|\gg t$. This can be understood by considering the Baker-Campbell-Hausdorff formula for the expansion of $W(x,t)=e^{iHt} W(x) e^{-iHt}$,
\be
W(x,t) = \sum_{k=0}^\infty \frac{(it)^k}{k!}\underbrace{[H, \dots[H}_k,W(x)] \dots], \label{eq:BCH}
\ee
where $H$ is the Hamiltonian which is assumed to consist of bounded local terms. As long as there is some sequence of terms in $H$ that connect the origin and point $x$ (and absent any special cancellations), the operator $W(x,t)$ will generically fail to commute with $V(0)$.

This does not necessarily imply that the magnitude commutator $[W(x,t), V(0)]$ between distance operators must be large. A bound of Lieb and Robinson \cite{Lieb:1972wy}, along with many subsequent improvements \cite{Nachtergaele2006,Hastings:2005pr,Hastings:2010loc}, limits the size of commutators of local operators separated in space and time, even in non-relativistic systems. In terms of the Heisenberg operator $W(x,t)$ at position $x$ and time $t$ and an operator $V$ at the origin of space and time, the bound reads
\beq
\| [W(x,t),V(0)]\| \leq K_0 \|W \| \|V\| e^{-(|x| - v_{\text{LR}} t)/\xi_0} \label{LR-bound}
\eeq
where $K_0$ and $\xi_0$ are constants, $\| \cdot \|$ indicates the operator norm, and $v_{LR}$ is the Lieb-Robinson velocity. The growth of the commutator is controlled by $v_{LR}$, which is a function of the parameters of the Hamiltonian. Hence although operators separated by a distance $x$ may cease to exactly commute for any $t> 0$, the Lieb-Robinson bound implies that their commutator cannot be $O(1)$ until  $t \gtrsim x / v_{LR}$.

Thus, the Lieb-Robinson velocity provides a natural notion of a ``light'' cone for non-relativistic systems. Even for relativistic systems, where causality implies that the commutator of local operators must be exactly zero for $t < x$,\footnote{In this letter we have set the speed of light to unity, $c=1$.} the ``Lieb-Robinson'' cone, if more restrictive, determines where in spacetime acting with a early operator $V(0)$ can nontrivially effect a later operator $W(x,t)$.

Commutators $[W(x,t), V(0)]$ of local Hermitian operators separated by time and space can also be used to characterize the butterfly effect in many-body quantum systems \cite{Larkin:1969abc,Almheiri:2013hfa}. The butterfly effect is naturally characterized in terms of such a commutator, which expresses the dependence of later measurements of distant operators $W(x,t)$ on an earlier perturbation $V(0)$. This connection was recently made sharp within the AdS/CFT correspondence, where quantum chaos in strongly-coupled large-$N$ gauge theories was shown to be connected to universal properties of high-energy scattering in the vicinity of the horizons of black holes \cite{Shenker:2013pqa}. This has inspired a large body of additional work \cite{Shenker:2013yza,Leichenauer:2014nxa,Roberts:2014isa,Kitaev:2014t1,Roberts:2014ifa,Shenker:2014cwa,Maldacena:2015waa,Kitaev:2014t2,Polchinski:2015cea,Marolf:2015jha,Hosur:2015ylk,Stanford:2015owe,Fitzpatrick:2016thx,Polchinski:2016xgd,Michel:2016kwn,Gu:2016hoy,Caputa:2016tgt,Swingle:2016var,Perlmutter:2016pkf,Sircar:2016old,Blake:2016wvh,Maldacena:2016chaos,Kitaev:2016chaos}.

To study the typical matrix elements of $[W(x,t), V(0)]$, it is useful to consider the average of its square,
\be
C(x,t) \equiv -\T{[W(x,t), V(0)]^2}, \label{eq:butterfly-commutator}
\ee
where $\T{\cdot}$ indicates thermal average at inverse temperature $\beta = 1/T$. $C(x,t)$ characterizes the strength of the butterfly effect at $x$ at time $t$ after an earlier perturbation of $V(0)$ at the origin. The statement of many-body chaos is that such commutators should grow to be large for almost all choices of operators $W,V$ \cite{Kitaev:2014t1,Roberts:2014ifa} and should remain large for a long time thereafter. The time when the commutator grows to be $O(1)$ (for suitably normalized operators) is known as the ``scrambling'' time \cite{Hayden:2007cs,Sekino:2008he,Shenker:2013pqa} and is usually denoted $t_*$. For large-$N$ gauge theories with $O(N^2)$ degrees of freedom per site, the early-time approach to scrambling is governed by an exponential growth with time
\be
C(x,t) = \frac{K}{N^2}e^{\lambda_L (t - x/v_B)} + O(N^{-4}),
\ee
for some constants $K$, $\lambda_L$, and $v_B$ that can depend on the choice of operators in the commutator $W,V$.\footnote{It can be that $K=0$, in which case the early-time growth is governed by the $O(N^{-4})$ term.} The onset of chaos is characterized by the two quantities: $\lambda_L$, and $v_B$.

$\lambda_L$ has been called a ``Lyapunov'' exponent \cite{Kitaev:2014t1}---in analogy with classical chaos---and characterizes the growth of chaos in time.\footnote{In \cite{Stanford:2015owe} it is argued that the classical limit of $\lambda_L$ does not always map onto the classical definition of the Lyapunov exponent, hence our use of ``quotes.''} Maldacena, Shenker, and Stanford \cite{Maldacena:2015waa} showed that this exponent is bounded by the temperature, $\lambda_L \le 2\pi/\beta$, with conjectured saturation for systems with thermal states that have a large-$N$ holographic black holes description whose near-horizon region is well-described by Einstein gravity.

$v_B$ is a velocity---the ``butterfly'' velocity---and characterizes speed at which the small perturbation grows \cite{Shenker:2013pqa}. Considering the commutator \eqref{eq:butterfly-commutator} as quantifying the effect of the perturbation $V(0)$ on $W(x,t)$, one may understand the butterfly effect as the growth of the operator $V(0)$ under time evolution. The speed of the growth is characterized by $v_B$, and the commutator begins to increase when $t \approx x/v_B + \lambda_L^{-1} \log N^2$ \cite{Roberts:2014isa}. This defines an effective light cone for chaos, a ``butterfly'' cone, outside of which the system is not affected by the perturbation.

In this letter, we explore the relationship between the Lieb-Robinson bound and the butterfly effect. A similarity was first noticed in \cite{Roberts:2014isa}, where it was pointed out that $v_{LR}$ can be used to bound the rate of growth of operators. Here, we would more directly like to contrast $v_{LR}$ with $v_B$. We will argue that the butterfly velocity can play the role of a low energy Lieb-Robinson velocity.

To elaborate, the Lieb-Robinson holds for any local lattice model of spins with bounded norm interactions. The constants $K_0$, $\xi_0$, and $v_{LR}$ depend on the model, but the general conclusion that there exists an effective light cone does not. However, as a bound on the operator norm of the commutator, it has some important limitations. It requires that $W$ and $V$ have finite operator norm. Also, the constants appearing in the bound are microscopic; from the point of view of a low energy description of the physics in terms of an emergent quantum field theory, they are UV sensitive. The reason for these limitations is that the Lieb-Robinson bound is \emph{state independent}. One could hope that given further information about the state of the system a tighter bound might hold. Such a bound would constitute a bound on matrix elements of the commutator between states of interest---e.g. the butterfly commutator \eqref{eq:butterfly-commutator}---instead of a bound on the operator norm.

To that end, we compute the rate of growth of commutators of generic local operators \eqref{eq:butterfly-commutator} for a wide variety of holographic states of matter. We show that butterfly commutators \eqref{eq:butterfly-commutator} grow ballistically with a butterfly velocity $v_B$, which is UV insensitive and only depends on IR quantities such as temperature and certain thermodynamic exponents. We use these results to argue that the butterfly velocity is a \emph{state-dependent} effective Lieb-Robinson velocity, which can be used to bound the growth of commutators, or rather their low energy matrix elements. For comparison, we provide a direct calculation of these butterfly commutators in a free fermion system. To further support our argument, we show that, if we are allowed only low energy operators, the butterfly velocity places an upper bound on the speed with which signals can be sent between distant parties. Finally, we briefly discuss the prospect that our results can probed experimentally using a recently proposed framework for measuring the scrambling time and the butterfly velocity \cite{Swingle:2016var}.

The details of our holographic calculations can be found in Appendix A and B, the free fermion calculation is stored in Appendix C, and the details of the signaling argument are in Appendix D.

\sbreak

Chaos in holographic models has so far been studied mostly in the special case of conformal field theories. Here, we will compute the rate of growth of commutators of local operators for a much wider class of holographic theories, specifically those with a finite density of charge for some conserved $U(1)$ symmetry.

The holographic backgrounds we study are solutions of Einstein-Maxwell-Dilaton (EMD) theory and describe the dynamics of a metric coupled to a gauge field and an uncharged scalar. With an electric flux for the gauge field turned on, such holographic models are dual to quantum field theories at finite density, that is field theories with a conserved $U(1)$ charge perturbed by a chemical potential.\footnote{Strictly speaking only some subset of these models have known string theory embeddings, e.g. \cite{Dong:2012se}. The remaining models appear to be consistent gravitational theories at low energies, but their UV status is not clear.} These models were originally studied in attempt to describe non-Fermi liquid states of electrons within the so-called AdS/CMT program \cite{Iizuka:2011hg,2012JHEP...01..125O,Swingle:2011np,Dong:2012se}.

The models are characterized by the bulk action of the EMD theory; the important information therein is the potential energy of the dilaton and the coupling of the dilaton to the field strength of the gauge field. In terms of observable parameters the backgrounds are characterized by two ``critical exponents,'' $z$ and $\theta$. The \emph{dynamical exponent} $z$ relates momentum to energy via $\omega \sim k^z$. Correlations obey power laws at zero temperature, but at finite temperature the field theory develops a correlation length given by $\xi \sim T^{-1/z}$. The exponent $\theta$ enters via the thermal entropy density,
\beq
s(T) \sim T^{(d-\theta)/z} \sim \(\frac{1}{\xi(T)}\)^{d-\theta},
\eeq
where $d$ is the spatial dimension of the field theory. The physics is this: at a conventional quantum critical point the entropy would scale as the inverse thermal length $\xi^{-1}$ to the power $d$, but when $\theta >0$ the entropy density scales like $\xi^{-1}$ to the power $d-\theta $. ``Hyperscaling'' is violated, so $\theta$ is called the \emph{hyperscaling violation exponent}. For additional details about these geometries, see Appendix~A.

The simplest example of such a hyperscaling violating theory is a Fermi gas at finite density. This system has $z=1$ and $\theta = d-1$. Since the fermions fill up their single particle energy levels up to a Fermi energy equal to the chemical potential, the locus of zero energy states in momentum space is generically $d-1$ dimensional instead of zero dimensional. The extent to which the holographic backgrounds we consider can describe conventional electronic states remains a topic of research.\footnote{For instance, note that imposing the null energy condition \eqref{null-energy} forbids $z=1$ and $\theta = d-1$ in holographic models.} However, since the results we find in the holographic model depend only on $z$ and $\theta$ in a rather simple way, we conjecture that they are more broadly applicable as we elaborate on below.

To calculate the growth of the commutator $C(x,t)$, we will study black holes geometries perturbed by a localized operator $W(x,t)$ \cite{Shenker:2013pqa,Roberts:2014isa}. For large $t$ such that $G_N e^{2\pi t / \beta} \sim 1$ (where $G_N$ is Newton's constant) backreaction will become important, and the perturbation will create a shock wave with a profile $h(x,t)$. Considering  $C(x,t)$ in the $t=0$ frame, the difference between the state created by $W(x,t) V(0)$  and the state created by $V(0) W(x,t)$ is a  null shift of the $V(0)$ quanta by $h(x,t)$ due to the shock wave \cite{Shenker:2013yza,Roberts:2014isa} (see also \cite{tHooft:1990fr,Kiem:1995iy}). Taking into account that the commutator is determined by the real part of $\T{W(x,t) \, V(0) \, W(x,t) \, V(0)}$ \cite{Roberts:2014isa}, $C(x,t)$ behaves at early times such that $\beta < t < x/v_B + t_*$ as
\be
C(x,t) \sim h(x,t)^2 + \dots,
\ee
where the scrambling time is given by $t_* = \frac{\beta}{2\pi} \log \frac{\lads^{d}}{G_N}$. We emphasize that this calculation is a statement about commutators of generic operators. The calculation is generic because of the universal coupling of energy to gravity: any operator that adds energy can backreact and generate a shockwave (and operators that don't add energy must commute with $H$ and don't scramble).

In Appendix~B, we present the technical details of our calculation of $h(x,t)$ using the shock wave techniques of \cite{Shenker:2013pqa,Roberts:2014isa}. From this, we extract the butterfly velocity for hyperscaling violating geometries
\be
v_B =  \bigg(\frac{\beta_0 }{\beta}\bigg)^{1-1/z} \sqrt{\frac{d+z-\theta}{2(d-\theta)}}, \label{hyperscaling-butterfly-velocity}
\ee
with $1/\beta_0$ a cutoff temperature (see Appendix~A) above which the holographic solution breaks down.
This is the main result of our letter. As promised, the butterfly velocity is a function of IR quantities: thermodynamic exponents $z, \theta$, and the inverse temperature $\beta$. The only dependence on high energy physics is the scale $\beta_0$, which simply sets the units of the temperature;\footnote{Suppose a hyperscaling violation metric of the form $ds^2 = r^{2\theta/d}(- A r^{2-2z} dT^2  + B dX^2 + C dr^2 )/r^2$ (with constants $A$, $B$, and $C$) is the IR limit of a solution which is asymptotically AdS (with metric $ds^2 = D\, (-dT^2 + dX^2 + dr^2)/r^2$ and constant $D$). Then, the change of variables $X = (C/B)^{1/2}\, x$ and $T = (C/B)^{1/2}\, t$ preserves the speed of light $c=1$ and puts the metric into the standard hyperscaling violating form \eqref{hyperscaling-metric} with $ r_0^{2z-2} = A/B$.} the $\beta_0$ dependence can be canceled by taking an appropriate ratio, so the prefactor is meaningful even when $z \neq 1$.\footnote{D.R. thanks Alexei Kitaev for raising this question.} Various comments are in order.

First, we note that in the limit of AdS $z=1,\theta = 0$, we recover the previously reported velocity for Einstein gravity \cite{Shenker:2013pqa,Roberts:2014isa}
\be
v_B(z=1,\theta =0) = \sqrt{\frac{d+1}{2d}}.
\ee
On the other hand, for the $z$ and $\theta$ appropriate to a Fermi gas at finite density, our result predicts
\be
v_B(z=1,\theta =d-1) = 1,
\ee
as appropriate for a theory that lives in effectively $1+1$-dimensions.

Second, we note that for $z<1$ the butterfly velocity diverges at $T=0$, but microscopic causality or even just a finite Lieb-Robinson velocity requires bounded $v_B$. Thus, if the hyperscaling violation geometry is to describe the deep IR, we require $z\ge1$. This means that $v_B$ has a temperature dependence of
\be
v_B \sim T^{1- 1/z}, \qquad  1- 1/z > 0,
\ee
and increases with temperature. $v_B$ behaves as a effectively ``renormalized'' Lieb-Robinson velocity, with a magnitude that depends on a negative power of the thermal scale $\beta$.

Third, the allowed values of $z$ and $\theta$ are constrained by the null energy conditions \eqref{null-energy}. For example, when $z=1$ the null energy condition implies that $\theta < 0$ or $\theta > d$. Furthermore, if $z\geq 1$ then $d - \theta + z \geq 0$. In fact, the stronger condition $d-\theta +z > 1$ follows from finiteness of energy fluctuations in the ground state \cite{Swingle:2015ipa}.

Finally, as a byproduct of our computation, we were able to show that for hyperscaling violating theories
\be
\lambda_L = \frac{2\pi}{\beta},
\ee
which, unlike $v_B$, is unchanged from its value in Einstein gravity \cite{Shenker:2013pqa}. We also point out that $\lambda_L$ behaves like a \emph{state-dependent} effective Lieb-Robinson growth rate (equivalent to $v_{LR} / \xi_0$ in \eqref{LR-bound}), a quantity for which a stronger bound was successfully derived \cite{Maldacena:2015waa}.

\sbreak

The main result of this letter is a computation of the butterfly velocity in a class of strongly coupled quantum field theories dual, via AdS/CFT, to Einstein-Maxwell-Dilaton theories of gravity. In fact, the computation relied only on the form of the metric, so any set of matter fields coupled to Einstein gravity which produces such a solution is sufficient. We framed the calculation in terms of the possibility of a bound on the growth of commutators analogous to the Lieb-Robinson bound in lattice many-body systems, so we now briefly discuss the thesis that the butterfly velocity functions as a low-energy Lieb-Robinson velocity.

To begin, it is interesting to compare the strongly coupled holographic results to results in the opposite limit of zero coupling. In Appendix~C, we record calculations of commutators for free particle lattice models. In these free particle models Wick's theorem implies that all commutators of composite operators are controlled by the basic commutator (or anti-commutator) between the elementary bosons (or fermions). For example, a one-dimensional free fermion hopping model with hopping matrix element $w$ is known to flow to a free CFT at low energies, and the anti-commutator of two fermions at integer positions $x$ and $y$ is
\be
\{c(x,t),c^\dagger(y,0)\} = e^{- i \pi (x-y)/2} J_{x-y}(2 w t), \label{main-text-anti-com}
\ee
where $J_\nu(x)$ is a Bessel function of index $\nu$. This anti-commutator grows outward like a shell,
\begin{align}
\{c(x,t),c^\dagger(y,0)\} &\sim \frac{t^{|x-y|} }{|x-y|!},  &&\mathrm{(early~times)}, \\
 &\sim t^{-1/2}, &&\mathrm{(late~times)},
\end{align}
showing an initial rise followed by a slow decay to zero. This is to be contrasted to a strongly-coupled system, where commutators should grow like a ball \cite{Roberts:2014isa}.

The anti-commutator \eqref{main-text-anti-com} is state-independent and sensitive to the UV details of the lattice model.\footnote{While naively this violates the bound of \cite{Maldacena:2015waa}, we note that the assumptions of the proof fail to hold: the relevant time-ordered four point functions fail to factorize after a ``dissipation time."}  However, by considering fermion operators corresponding to low energy wavepackets, it is possible to exhibit an anti-commutator which grows instead at some group velocity given by the momentum derivative of the energy evaluated at the average momentum of the wavepacket. If the dispersion relation near zero energy goes like $\epsilon \sim k^z$ with $k$ the momentum, then in a thermal state where the typical energy is $T$, the typical thermal group velocity will be $T^{1 - 1/z}$ as in the holographic results. With the additional reasonable assumption that weak interactions cause high energy particles to decay towards low energy, it is plausible that our holographic result applies at both weak and strong coupling, at least in terms of its temperature dependence. Note also that this result is not a trivial consequence of dimensional analysis when $\theta \neq 0$ since there are additional scales in the problem.

In addition to these statements about commutators, there is an alternative interpretation of $v_B$ in terms of quantum information flow. As we show in Appendix~D, within a certain model of communication using only low energy observables, if the relevant commutators between operators controlled by two distant parties are small, then these parties cannot communicate much information. The butterfly velocity thus defines an information theoretic light-cone in which quantum information cannot be spread in space faster than $v_B$. Relatedly, in \cite{Hosur:2015ylk} it was argued that $v_B$ controls the flow of information through time, and in \cite{Mezei:2016ent}, it will be shown that the butterfly velocity $v_B$ controls the rate of growth of the entanglement wedge in holographic theories. Thus, $v_B$ is both a measure of the growth of operators / commutators and also a measure of the spreading of information under time evolution. These results all support our identification of the butterfly velocity as a low-energy Lieb-Robinson velocity.

It is also interesting to compare $v_B$ with the speed of hydrodynamic sound. For conformal field theories it is well known that there exists a hydrodynamic sound mode with velocity $v_s = \frac{1}{\sqrt{d}}$. Such theories correspond to the special case $z=1$ and $\theta = 0$, where the butterfly velocity is $v_B = \sqrt{\frac{d+1}{2d}}$. As necessary for $v_B$ to be an upper bound on operator growth, we find $v_B \geq v_s$ for CFT. When $z > 1$, our results provide an interesting constraint on the speed of the sound: if a hydrodynamic sound mode exists, its velocity must go to zero with temperature at least as fast as $v_B$.\footnote{For example, a weakly interacting thermal gas of particles with dispersion $\omega \sim k^2$, i.e. non-relativistic particles, may be approximately described as an ideal gas with sound speed $\propto T^{1/2}$.}$^,$\footnote{Note: in the ground state, scrambling can still occur but might be slower \cite{Roberts:2014ifa}. For example, consider a $2d$ large-$c$ CFT on an infinite line. In this case, the commutator will vanish outside the butterfly cone due to causality (since $v_B=1$), but then it will begin to grow only as a power law $~(t-x)^4$.} This would be interesting to check using holography. The constraint can be avoided by adding extra ingredients into the bulk, e.g. probe D-branes \cite{2012JHEP...11..028A}. Presumably a proper calculation of the butterfly velocity in the presence of the D-brane would be sensitive to any sound mode on the D-brane world volume.

As a final point, we discuss the relationship between the butterfly velocity and the so-called ``entanglement velocity'' $v_E$ \cite{Liu:2013iza,Hartman:2013qma}. This velocity is relevant for the following physical setup: a rapid injection of energy into a system which then thermalizes at some final equilibrium temperature $T$ (depending on the amount of energy injected) with a corresponding coarse-grained thermal entropy density $s$. During this process the entanglement entropy of a region $R$ with boundary size $|\partial R|$ changes, and $v_E$ is defined by the equation \cite{Liu:2013iza,Hartman:2013qma}
\beq
\frac{d S(R)}{dt} = s \, |\partial R| \,v_E.
\eeq

$v_E$ can be estimated as follows. In terms of the thermal length $\xi$, we divide the boundary $\partial R$ into thermal cells of size $\xi^d$. Each thermal cell is naturally associated with a thermal rate $T$ and contains a number of active degrees given by $s \xi^d$. Since entanglement is only generated at the boundary, the amount of entanglement generated per unit time is roughly the size of $\partial R$ in units of the thermal length times the number of degrees of freedom per thermal cell times the thermal rate, i.e.
\beq
\frac{dS(R)}{dt} \sim \frac{|\partial R|}{\xi^{d-1}} \times s \xi^d \times T = s |\partial A| \xi T.
\eeq
This gives the estimate $v_E \sim \xi T$ or $v_E \sim T^{1- 1/z}$ up to a $z$ and $\theta$ dependent prefactor. (We have calculated the entanglement velocity holographically using the methods of \cite{Hartman:2013qma} and find agreement with the scaling estimate; for a complete analysis see \cite{2014PhRvD..90d6004A,2014JHEP...08..051F}.) However, while $v_E$ and $v_B$ scale similarly with temperature, it is not clear that $v_E$ is directly related to the spread of quantum information.\footnote{It was observed for holographic CFTs that $v_B$ is equal to so-called saturation velocity in the entanglement growth result ($v_E$ is relevant at intermediate times before the entropy saturates). This also holds in our case \cite{2014PhRvD..90d6004A,2014JHEP...08..051F}. This is general property of holographic theories described by Einstein gravity \cite{Mezei:2016ent}.} It may be better thought of as a rate (times a thermal length), although it has been shown to be bounded by unity in relativistic theories \cite{Casini:2015zua,Hartman:2015apr}.

\sbreak

To summarize, we have seen that the butterfly velocity $v_B$ characterizes the spread of quantum information in quantum many-body systems and can provide a low-energy analog of the microscopic Lieb-Robinson velocity $v_{LR}$. We have also computed the dependence of the butterfly velocity on temperature and on the thermodynamic exponents $z$ and $\theta$ at weak and strong coupling. Given the interest in experimental measures of the spread of quantum information \cite{Cheneau12,Richerme14,Jurcevic14,Eisert15}, it is natural to ask if our results can be tested experimentally.

Recently, an experimental protocol was devised to measure out-of-time-order correlation functions \cite{Swingle:2016var}. Such correlators are simply related to the squared commutator $C(x,t)$, and therefore these measurements are necessary to gain access to the butterfly velocity. Furthermore, although the simpler measurement of an average commutator $\langle[W(x,t),V(0)]\rangle$ would give information about the butterfly velocity (at least a lower bound), a measurement of $C(x,t)$ is more desirable because it gives information about typical off-diagonal matrix elements.

Cold bosonic atoms moving in an optical lattice constitute one experimental setting where such measurements could be performed. Because the sign of the Hamiltonian can be effectively reversed by combining lattice modulation with Feshbach resonance, the echo-like measurement necessary to measure $C(x,t)$ is conceivable.

Furthermore, the model can be realized at a variety of particle densities. For example, at low density the system is described by a $z=2$ theory, while at commensurate densities the system can be tuned to a quantum critical point described by the $XY$ CFT. The same model but with fermionic instead of bosonic atoms can realize a Fermi liquid with tunable interactions, a system with non-zero $\theta$. It would be interesting to see whether the result \eqref{hyperscaling-butterfly-velocity} persists in these systems.

{\bf{Note Added:}}
After this letter was completed \cite{Blake:2016wvh} appeared which also computes the butterfly velocity in hyperscaling violating geometries.

\section*{Acknowledgments}
We thank Mark Mezei and Douglas Stanford for discussions and Adam Brown, Leonard Susskind, and Ying Zhao for collaboration on related work.
DR is supported by the Fannie and John Hertz Foundation and also acknowledges the U.S. Department of Energy under cooperative research agreement Contract Number DE-SC0012567.
BGS is supported by the Simons Foundation and the Stanford Institute for Theoretical Physics. This paper was brought to you by upper bounds and chaos.

\section{Appendix A: Hyperscaling violating black holes}

The $d+2$-dimensional metric of the black hole solution in the hyperscaling violating theory in planar-Schwarzschild coordinates takes the form \cite{Iizuka:2011hg,2012JHEP...01..125O,Swingle:2011np,Dong:2012se}
\be
ds^2 = \lads^2 \, \bigg(\frac{r}{r_0}\bigg)^{2 \theta  / d} \Big[ -f(r) r_0^{2z-2} \frac{dt^2}{r^{2z} } +\frac{1}{f(r)} \frac{dr^2}{r^2} + \frac{dx^i dx^i}{r^2} \Big], \label{hyperscaling-metric}
\ee
with dynamical exponent $z$, hyperscaling violating exponent $\theta$, spacetime curvature radius $\lads$, and a dynamical scale $r_0$. Here, $i$ runs over the $d$ transverse dimensions. The boundary lives at $r=0$, and the solution \eqref{hyperscaling-metric} is thought to be a good description for $r \sim r_0$.

The solution \eqref{hyperscaling-metric} is characterized by the function $f(r)$
\be
f(r) = 1 - \bigg( \frac{r}{r_h} \bigg)^{d-\theta + z}, \label{emblack}
\ee
with the black hole horizon at $r_h$, and the requirement $r_0 < r_h$. The inverse temperature is
\be
\beta = \frac{4\pi}{|f'(r_h)|}\bigg( \frac{r_h}{r_0} \bigg)^{z-1}  = \frac{4 \pi r_h}{|d-\theta+z|}  \bigg( \frac{r_h}{r_0} \bigg)^{z-1}, \label{temperature}
\ee
and the null energy condition constrains the parameters as
\begin{align}
&(d-\theta)(d(z-1) - \theta) \ge 0, \label{null-energy}\\
&(z-1)(d+z-\theta) \ge 0. \notag
\end{align}
Combined with the assumption $z>1$, this fixes $d-\theta+z > 0$ (and henceforth we will drop the absolute value sign).
Finally, when matching onto an AdS geometry at small $r$, the radius $r_0$ marks the location where the IR solution fails. The condition $r_0 < r_h$ translates into a cutoff temperature $T_0 = 1/\beta_{0}$
\be
\beta_{0} =  \frac{4 \pi r_0}{d-\theta+z}, \label{max-temp}
\ee
above which \eqref{hyperscaling-metric} is not a good description. Finally, the entropy density of the state is
given by
\beq
s = \frac{\ell^d}{4G} \left(\frac{r_h}{r_0}\right)^\theta \left(\frac{1}{r_h}\right)^d,
\eeq
and using \eqref{temperature} we see
\beq
s(T) \sim T^{(d-\theta)/z}.
\eeq

\section{Appendix B: Holographic calculation of the butterfly velocity}
In this appendix, we will present a general calculation of the butterfly velocity $v_B$ for spacetimes with matter. While this calculation closely follows \cite{Roberts:2014isa}, there is a subtly for theories coupled to matter \cite{Sfetsos:1994xa}. Therefore, we will present the calculation in complete generality before specializing to the hyperscaling violating geometries \eqref{hyperscaling-metric}.

Consider a $d+2$-dimensional spacetime holographically dual to a $d+1$-dimensional field theory in a thermal state, with $d$ planar spatial dimensions. Let us assume the metric can be put in Kruskal form
\be
ds^2 = 2 A(uv)\, du dv + B(uv) \, dx^i dx^i, \label{unperturbed-metric}
\ee
where $i$ runs over the $d$ transverse directions. The horizon is at $uv = 0$, where $A(0)$, $B(0)$, and $B'(0)$ are all smooth and nonzero. In these coordinates, boundary is at $uv=-1$, and the black hole singularity is at $uv=1$. These coordinates cover the maximally extended eternal black solution, with two entangled boundaries connected by a wormhole \cite{Maldacena:2001kr}. In our conventions, Schwarzschild time on the left boundary $t_L$ runs backwards. Our goal is to perturb this spacetime by acting with a local, thermal scale precursor operator on the left boundary at $x=0$ and time $t_L = t$. The butterfly velocity is given the rate of growth of this perturbation in transverse directions.

For large $t$, we can represent the perturbation by a stress tensor localized on the $u=0$ horizon
\be
T_{uu}^{shock} = \frac{P}{\lads^{d+2}}e^{\frac{2\pi}{\beta} t} \delta(u) \delta(x),
\ee
where $P$ is related to the asymptotic momentum of the perturbation, and the delta function is $x$ is only really valid for $|x|>\beta$. As explained in \cite{Sfetsos:1994xa}, the matching condition between the geometries along the shock also requires considering additional terms for theories coupled to matter. If our original metric \eqref{unperturbed-metric} was a solution to Einstein's equations with stress tensor $T_{\mu\nu}^0$, then in the geometry with the shock wave we have to add $T^1_{\mu\nu}$, with
\begin{align}
T^1_{uu} &= \frac{P}{\lads^{d+2}}e^{\frac{2\pi}{\beta} t} \delta(u) \delta(x)  - 2 h(x,t) \delta (u) \, T^0_{uv}, \\
T^1_{uv} &=  -  h(x,t) \delta (u) \, T^0_{vv},
\end{align}
and all other components vanishing. For large $t$, this perturbation will backreact the geometry to give a shock wave solution
\be
ds^2 = 2 A(uv)\, du dv + B(uv) \, dx^i dx^i -2 A(uv) h(x,t) \delta(u) \, du^2. \label{metric}
\ee
Plugging into Einstein's equations,
\be
R_{\mu\nu} - \frac{1}{2} g_{\mu\nu} R = 8 \pi G_N (T^0_{\mu \nu} + T^1_{\mu \nu}),
\ee
we find that we have a solution so long as this differential equation is satisfied
\be
(\partial_i \partial_i  - \mu^2) h(x,t) = 8\pi G_N\frac{B(0)}{A(0)}\frac{ P}{\lads^{d+2}} e^{\frac{2\pi}{\beta} t}   \delta(x),
\ee
with $\mu^2$ given in terms of $A(0)$, $B'(0)$, and $d$ as
\be
\mu^2 = \frac{d}{2} \frac{B'(0)}{A(0)}.
\ee
For large distance $|x|$ and $P$ thermal scale, the solution is given by
\be
h(x,t) =  \frac{e^{ \frac{2\pi}{\beta} (t - t_*) - \mu |x| } }{|x|^{ \frac{d-1}{2} }},
\ee
with the fast scrambling time \cite{Hayden:2007cs,Sekino:2008he,Shenker:2013pqa} $t_* \approx \frac{\beta}{2\pi} \log N^2$.\footnote{We note in passing that for all metrics of this form, the exponent $\lambda_L$ is equal to its Einstein value: $\lambda_L = 2\pi / \beta$. } The function $h(x,t)$ is the (position dependent) shift induced by crossing the shock wave at $u=0$. It characterizes the growth of the perturbing operator in the boundary theory. The butterfly effect velocity $v_B$ is given by the combination
\be
v_B = \frac{2\pi}{\beta \mu}. \label{buttefly-velocity}
\ee

Now, we will specialize to hyperscaling violating metrics \eqref{hyperscaling-metric}. To find $v_B$, we need to express it in Kruskal form \eqref{unperturbed-metric}. The tortoise coordinate defined by the relation
\be
dr_* = -  \frac{r^{z-1}}{r_0^{z-1}}  \frac{dr}{f(r)}, \label{tortoise}
\ee
and we can define Kruskal coordinates
\be
u v = - e^{4 \pi a r_*(r) / \beta}, \qquad u/v = - e^{- 4\pi t / \beta }, \label{kruskal-def}
\ee
with $a \equiv (r_h / r_0)^{z-1}$.
Ultimately, we find
\begin{align}
&A(uv)=-\frac{2f(r)}{f'(r_h)^2}\frac{\lads^2 r_h^{2z-2}}{r^{2z} } \bigg(\frac{r}{r_0}\bigg)^{2\theta/d}  e^{- 4 \pi r_*(r) / \beta} \notag,\\
&B(uv) =\frac{\lads^2}{r^2} \bigg(\frac{r}{r_0}\bigg)^{2\theta/d}. \label{A-and-B-def}
\end{align}

To find $v_B$, we need to evaluate $A(0)$  and $B'(0)$. Expanding $A(uv)$ near the horizon, using the fact that
\begin{align}
&f(r) \approx (r_h - r) f'(r_h) + \dots, \notag \\
&e^{ 4 \pi r_*(r) / \beta} \approx (r_h - r) \kappa + \dots,
\end{align}
for some constant $\kappa>0$, we find
\be
A(0) = - \frac{ 2\lads^2  }{ \kappa  f'(r_h) r_h^{2}} \bigg(\frac{r_h}{r_0}\bigg)^{2\theta/d}. \label{A0}
\ee
To get $B'(0)$, we differentiate \eqref{A-and-B-def} to find
\be
B'(0) =r'(0)\frac{\lads^2}{r_h^3} \bigg(\frac{2\theta}{d} -2\bigg)\bigg(\frac{r_h}{r_0}\bigg)^{2\theta/d}, \label{B-prime-0}
\ee
and using \eqref{tortoise} and \eqref{kruskal-def}, we evaluate
\be
r'(0) = \frac{dr}{dr_*} \frac{dr_*(uv)}{d(uv)}\bigg|_{uv=0} = \frac{1}{\kappa}.
\ee
Using \eqref{A0} and \eqref{B-prime-0}, we can evaluate the butterfly velocity for hyperscaling violating geometries expressed in terms of inverse temperature $\beta$, which gives our result \eqref{hyperscaling-butterfly-velocity}.

\section{Appendix C: Free fermion calculations}

In this appendix, we evaluate many of the quantities considered in this letter for free particles. We focus on fermions, but a completely analogous set of calculations are possible for bosons.

Consider a set of fermion annihilation operators $c(x)$ labeled by position $x$ and obeying the algebra $\{c(x),c^\dagger(y)\} = \delta_{x,y}$. The natural role of the commutator is here played by the anti-commutator
\be
A(x,t) = \langle \{ c(x,t), c^\dagger(0,0)\} \rangle. \label{def:anti-com}
\ee
Assuming the dynamics is given by a quadratic Hamiltonian, then the anti-commutator is a state independent for all $x,t$ and proportional to the identity.\footnote{Note that since
\be
\langle \{c(x,t),c^\dagger(0)\}^\dagger \{c(x,t),c^\dagger(0)\} \rangle = |A(x,t)|^2,
\ee
it is not necessary to separately consider the squared anti-commutator.}

Let the quadratic Hamiltonian be
\beq
H = \sum_k \epsilon_k \, c_k^\dagger c_k
\eeq
where $k$ labels the momentum. Writing the mode expansion of $c(x)$ in terms of energy modes as
\beq
c(x) = \sum_k \phi_k(x) c_k,
\eeq
then anti-commutator \eqref{def:anti-com} is
\beq
A(x,t) = \sum_{k} \phi_k(x) \phi_{k}(0)^* e^{-i \epsilon_k t}, \label{free-anti-com}
\eeq
which is independent of state as claimed. This is also the commutator of bosonic creation and annihilation operators when the bosonic dynamics is quadratic.

To be concrete, let's specialize to free fermions propagating in a one-dimensional wire of length $L$. The mode functions are
\beq
\phi_k(x) = \frac{e^{-i k x}}{\sqrt{L}},
\eeq
and the energy is given by $\epsilon_k = - 2 w \cos k$,  where $w$ is the nearest neighbor hopping matrix element such that in the position basis the Hamiltonian is
\be
H = -w \sum_{x} c(x)^\dagger c(x+1) + \textrm{h.c.}
\ee
After plugging into \eqref{free-anti-com}, writing the sum as an integral, shifting integration variables to $q = k - \pi/2$, and using $\cos k = \cos(q+ \pi/2) = - \sin q$, we finally obtain
\beq
A(x,t)  = e^{i \pi x/2} \int_0^{2\pi} \frac{dq}{2\pi} e^{i [xq - 2 w t \sin q]},
\eeq
where $x$ is the positive distance between the operators.
This quantity is a standard integral representation of a Bessel function, so the anti-commutator can be written as
\beq
A(x,t) =  e^{i \pi x/2} J_{x}(2wt).
\eeq
When the index of the Bessel function is much larger than its argument, the Bessel function is approximately
\beq
J_{x}(2wt) \approx \frac{(wt)^{x}}{x!}. \label{fermion-anti-com}
\eeq
Thus, the growth of the free-fermion anti-commutator is power law like, with a power given by the (integer) separation between the sites of interest. This is the main result of this appendix.

Let us discuss some consequences of \eqref{fermion-anti-com}.  First, note that this quantity becomes of order one when $wt \sim x$, indicating a ballistic growth of the anti-commutator with speed set by the parameter $w$. Second, unlike the strongly-interacting many-body chaotic systems, at late times the anti-commutator decays to zero slowly like $t^{-1/2}$. The growth of a free particle is a ``shell'' rather than a ``ball.'' Third, since the anti-commutator is state independent, it is UV sensitive. To show this, let us consider a general dispersion relation $\epsilon_k$ and evaluate $A(x,t)$ in a stationary phase approximation. Then, defining a velocity $v_k \equiv \partial_k \epsilon_k$, we find that the points of stationary phase are given by those $k$ such that
\beq
\frac{x}{t} = v_k.
\eeq
If $v_k$ is bounded and if $x \gg t$, then there is no point of stationary phase. The point of stationary phase appears when
\beq
\frac{x}{t} = v_{\max} = \max_k v_k,
\eeq
which is manifestly UV sensitive quantity since the maximum is taken over all momenta.

However, when considering smeared operators it is possible to get a UV insensitive growth rate. Consider the smeared operator
\beq
c_w(x_0,k_0) = \sum_x \frac{e^{i k_0 x - \frac{(x-x_0)^2}{4 \sigma^2}}}{(2 \pi \sigma^2)^{1/4}} c(x)
\eeq
which corresponds to a properly normalized wavepacket with average position $x_0$ and average momentum $k_0$. The anti-commutator is defined
\beq
A_w(x_0,t) = \langle \{ c_w(x_0,k_0,t), c_w^\dagger(0,k_0,0)\} \rangle,
\eeq
where for simplicity both operators have the same average momentum. Henceforth the momentum label will be suppressed. In terms of the original anti-commutator $A$ the wavepacket anti-commutator is
\beq
A_w(x_0,t) = \sum_{x, y} \frac{e^{i k_0 x - \frac{(x-x_0)^2}{4 \sigma^2}}}{(2 \pi \sigma^2)^{1/4}} \frac{e^{-i k_0 y - \frac{y^2}{4 \sigma^2}}}{(2 \pi \sigma^2)^{1/4}} A(x-y,t).
\eeq
A short calculation shows that the wavepacket anti-commutator takes the form
\beq
A_w(x_0,t) \sim \frac{1}{\sqrt{\sigma^2 + i \gamma t}} \exp \Big\{ - \frac{(x_0 - v_{k_0} t)^2}{4 \sigma^2 + i \gamma t } \Big\} \label{smeared-A}
\eeq
where we have defined $\gamma = \partial_k^2 \epsilon_k |_{k_0}$, and the group velocity is $v_{k_0}$. Thus, we see the smeared anti-commutator travels with a velocity set by the average momentum $k_0$. Furthermore, the pattern of operator growth is clearly ``shell-like'' since at late times the magnitude of \eqref{smeared-A} decays like $t^{-1/2}$.

\sbreak

In the many-body fermion problem, the thermal state is described by the Fermi occupation numbers $n_k = (e^{\beta \epsilon_k} + 1)^{-1}$. At low temperature, the negative energy states are mostly filled and positive energy states are mostly empty. Suppose that near a zero of the energy the dispersion relation has the form $\epsilon_k \sim \alpha (k-k_F)^z$ where we allow for a general Fermi momentum $k_F$. Integer values of $z$ can be easily achieved with suitable band-structure engineering. The typical single particle excitation in the thermal state will have energy of order $T$, which corresponds to a momentum relative to the Fermi momentum of order $(k-k_F) \sim T^{1/z}$. The group velocity with this dispersion is $v_k = \alpha z (k-k_F)^{z-1}$, so setting $k - k_F = T^{1/z}$, we obtain a velocity in the anti-commutator of order
\beq
v_{k_0} \sim T^{1 - 1/z},
\eeq
as in the holographic computation \eqref{hyperscaling-butterfly-velocity}. Of course, the free fermion model has no scattering, so information can easily be carried by higher energy excitations at a much higher velocity. However, we speculate that in the presence of interactions such high energy quanta will rapidly decay down to the thermal scale where they will move with a velocity of order $T^{1-1/z}$.

Finally, let us note that although we focused on the anti-commutator of a single fermion mode, Wick's theorem (assuming the state is Gaussian) immediately implies that no other square commutator can grow substantially faster than the fermion anti-commutator. Roughly speaking, this is because if distant fermions perfectly anti-commute, then distant bosonic operators built from them will perfectly commute. Take for example the density $n(x,t) = c^\dagger(x,t) c(x,t)$. A quick use of the anti-commutation relations and Wick's theorem shows that both $\langle [n(x,t),n(0)] \rangle$ and $\langle [n(x,t),n(0)]^2 \rangle$ are proportional to $\{c(x,t), c^\dagger(0,0)\}$ (or its conjugate), although they are no longer state independent.

\section{Appendix D: Signaling speeds}

Even in models with microscopic Lorentz invariance, we have shown that the butterfly velocity can be much less than the speed of light at low energy. This can also be true for lattice models with a butterfly velocity that is smaller than the microscopic Lieb-Robinson velocity. This suggests that the butterfly velocity might provide a tighter bound on the speed at which information can propagate.

Consider two parties, $A$ and $B$, such that $B$ wishes to transmit a bit $b$ to $A$. $A$ and $B$ share parts of a general quantum state $|\psi\rangle_{ABE}$, where $E$ is the environment. $B$ can send information to $A$ by applying a local unitary to the state: $U_0$ to send $b=0$, or $U_1$ to send $b=1$. $A$ will try to determine $B$'s message using a measurement of $\{M_0,M_1\}$ with $M_0 + M_1 = I$. Finally, let us assume that the system evolves for a time of $t$  after $B$ applies the unitary and before $A$ makes the a measurement.

Given that $B$ sends message $b$ by applying $U_b$, the probability that $A$ obtains answer $a$ is
\beq
p(a|b) = \langle \psi | U_b^\dagger e^{i H t}  M_a e^{-i H t} U_b | \psi \rangle = \langle \psi | U_b^\dagger M_a(t) U_b | \psi\rangle
\eeq
where $M_a(t) = e^{i H t} M_a e^{-i H t}$ is a Heisenberg operator. A simple rearrangement gives
\beq
p(a|b) = \langle M_a(t) \rangle + \langle U_b^\dagger [U_b , M_a(t)]\rangle.
\eeq
Thus if $[U_b, M_a(t)] = 0$, then the conditional probability $p(a|b)$ is independent of $b$ and  no information can be sent. This is the usual notion of exact causality.

Suppose instead that the commutator is only small, bounded such that
\beq
\langle U_b^\dagger [U_b, M_a(t)]\rangle = \epsilon q(a,b),
\eeq
where $q$ is real and $\epsilon$ is chosen such that $|q| \leq 1$ for all relevant (i.e low-energy) choices of $U$ and $M$. The classical capacity of this channel is
\be
C_{AB} = \max_{p(b),U,M} I(A:B),
\ee
where $I(A:B)$ is the classical mutual information between $A$ and $B$. Using the Fannes inequality \cite{Fannes:1973cty}, we can bound the mutual information
\be
I(A:B) \le 2\sum_{ab}|p(a,b) - p(a)p(b)|.
\ee
Using the fact that $p(a) = \langle M_a(t) \rangle  + O(\epsilon)$, we see that the channel capacity must be
\be
C_{AB} = O(\epsilon),
\ee
and therefore the signaling speed is set by the growth of commutators.

Although the quantum capacity is harder to compute, we expect on general grounds that it will also be small. For example, with shared entanglement quantum information can be transmitted via teleportation, but this method also requires classical communication.

\bibliography{hyper}

\end{document}